\documentstyle{mn}

\input{epsf}

\def\lsim{\mathrel{\hbox{\rlap{\hbox{\lower4pt\hbox{$\sim$}}}\hbox{$<$}}}}
\def\gsim{\mathrel{\hbox{\rlap{\hbox{\lower4pt\hbox{$\sim$}}}\hbox{$>$}}}}

\def\and   {\rm {et al.} \rm}  

\begin{document}

\title
[
Higher Order Clustering in the Durham/UKST and Stromlo-APM 
Galaxy Redshift Surveys
]
{
Higher Order Clustering in the Durham/UKST and Stromlo-APM 
Galaxy Redshift Surveys
}

\author[F. Hoyle  et al. ]
{
Fiona Hoyle$^{1}$, 
Istvan Szapudi$^{1,2}$,
Carlton M. Baugh$^{1}$
\\
1. Department of Physics, Science Laboratories, South Road, Durham DH1 3LE \\
2. Present Address: CITA, University of Toronto, 60 George Street, 
Ontario, Canada, M5S 3H8, 
\\
\\
}

\maketitle 
\vspace{-1cm}
 
\begin{abstract}

We present a counts-in-cells analysis of clustering in the 
optically selected Durham/UKST and Stromlo-APM Galaxy Redshift Surveys. 
Minimum variance estimates of the second moment, skewness ($S_3$) and 
kurtosis ($S_4$) of the count probability distribution are 
extracted from a series of volume limited samples of varying  
radial depth. 
The corresponding theoretical error calculation takes into account 
all sources of statistical error on the measurement of the moments,
and is in good agreement with the 
dispersion over mock redshift catalogues.
The errors that we find on $S_3$ and $S_4$ are  
larger than those quoted in previous studies, in spite of 
the fact that the surveys we consider cover larger volumes.
$S_3$ varies little with cell size, with values in the range $1.8-2.2$ 
and errors $ \lsim20\%$, for cubical cells of side $3-20 h^{-1}$Mpc.
Direct measurements of $S_3$ are possible 
out to $\sim 35 h^{-1}$Mpc, though with larger errors.
A significant determination of $S_4$ is only possible for one scale, 
${\it l} \sim 6 h^{-1}$Mpc, with $S_4 \approx 5$.
We compare our results with theoretical predictions from  
{\it N}-body simulations of cold dark matter universes. 
Qualitatively, the skewness of the dark matter has the same form 
as that of the galaxies. 
However, the amplitude of the galaxy $S_3$ is lower than that 
predicted for the dark matter. 
Our measurements of $S_3$ are consistent with the predictions 
of a simple model in which initially Gaussian fluctuations in 
the dark matter evolve gravitationally, if a second order bias 
term is specified, in addition to the traditional linear bias, 
in order to describe the relation between the distribution of 
galaxies and dark matter. 
\end{abstract}


\section{INTRODUCTION}

Maps of the local universe have improved dramatically over the
last decade and permit the clustering pattern of galaxies
to be quantified on large scales (e.g. Efstathiou et al. 1990(a); 
Maddox et al. 1990; Saunders et al. 1991).
Such observations can potentially constrain both the nature of
the dark matter and the statistics of primordial density fluctuations.

The first accurate measurements of the galaxy two-point correlation
function on scales greater than $10h^{-1}$Mpc indicated more structure
than expected in the simplest form of the cold dark matter (CDM) model.
This led to variants of the CDM model being studied
(Efstathiou, Sutherland \& Maddox 1990).
Currently, the most successful CDM model is a low density,
spatially flat universe with a cosmological constant, $\Lambda$CDM. 
The power spectrum in the $\Lambda$CDM model is described
by a shape parameter $\Gamma = 0.2-0.3$ (in this Letter we use
the parameterisation of the power spectrum given in
Efstathiou, Bond \& White 1992).
If fluctuations in the dark matter are normalised so as to reproduce
the local abundance of hot X-ray clusters (White, Efstathiou \& Frenk 1993),
the power spectrum in the $\Lambda$CDM model is similar to that observed
for galaxies on scales around $k \sim 0.05 - 0.2 h {\rm Mpc}^{-1}$
(Gazta\~{n}aga \& Baugh 1998).
On small scales, however, when the effects of peculiar velocities
are ignored (real space), the dark matter power spectrum has a higher
amplitude than the galaxy power spectrum (Gazta\~{n}aga 1995; Peacock 1997;
Jenkins et al. 1998).
Furthermore, the small scale power spectrum for galaxies is a power law
over a decade and a half in wavenumber, whereas the dark matter correlation
function shows considerable curvature.

\begin{table*}
\begin{tabular}{lccccccc}
Survey & cell size & R$_{\rm max}$  & Volume & M$_{\rm crit}$ - 5log$h$ & N$_{\rm gal}$ & $S_3$ & $S_4$ \\
 & (h$^{-1}$Mpc) & ($h^{-1}$Mpc) & (10$^6$h$^{-3}$Mpc$^3$) &  & &  \\ \hline
Durham/UKST & 3.125 & 170 & 0.721 & -19.58 & 510 & 1.94$\pm$0.14 & 1.5 \\
Durham/UKST & 6.3125 & 180 & 0.855 & -19.73 & 515 & 2.11$\pm$0.08 & 5.0$\pm$3.8 \\
Durham/UKST & 12.625 & 180 & 0.855 & -19.73 & 515 & 1.82$\pm$0.21 & 3.0 \\
Durham/UKST & 25. & 170 & 0.721 & -19.58 &  510 & 1.67$\pm$1.32 & 2.2 \\ \hline
Stromlo-APM & 3.9375 & 180 & 2.547 & -19.45 & 471 & 2.07$\pm$0.57 & 13. \\
Stromlo-APM & 8.875  & 180 & 2.547 & -19.45 & 471 & 1.89$\pm$0.17 & 3.1 \\
Stromlo-APM & 18.1875 & 190 & 2.995 & -19.58 & 465 & 2.24$\pm$0.29 & 8.2 \\ 
Stromlo-APM &  36.625 & 200 & 3.493 & -19.71 & 434 & 1.41$\pm$1.01 & - \\ \hline
\end{tabular}
\caption{Minimum variance estimates of $S_3$ and $S_4$ in cubical cells from 
the Durham/UKST and the Stromlo-APM Surveys. The errors on $S_3$ are the 
$1 \sigma$ theoretical errors for a sample with the volume, 
geometry and number of galaxies used in the measurement.
The relative errors on the estimates of $S_4$ are greater 
than 100 per cent apart from one Durham/UKST value.}
\label{tab:best}
\end{table*}

Heuristic biasing schemes, in which the galaxy distribution is 
proposed to be a local transformation of the smoothed density field,
have enjoyed a certain degree of success in reproducing the observed 
correlation function (Coles 1993; Cole et al. 1998; Mann, Peacock 
\& Heavens 1998; Narayanan et al. 1999). 
Progress towards a physical understanding of the processes
responsible for producing a bias between the galaxy and
dark matter distributions has been made using semi-analytic
models for galaxy formation (Benson et al. 2000a,b; Kauffmann et al. 1999).
In a $\Lambda$CDM model that reproduces the bright end of the
field galaxy luminosity function, Benson et al. find remarkably good 
agreement with both the amplitude and power law slope of the correlation 
function of APM Survey galaxies (Baugh 1996).
If the distortions to the clustering pattern caused by peculiar motions 
are included, the correlation function of the dark matter is very similar
to that of the semi-analytic galaxies in the $\Lambda$CDM model,
with no bias seen on small scales.
The correlation function is also in good agreement with the measurements 
from galaxy redshift surveys (cf Fig. 1 of Benson et al 2000b).

The constraints on models of galaxy formation provided by the
two-point correlation function are somewhat limited.
The second moment gives a full statistical description of
the density field only in the case of very weak fluctuations.
Galaxy clustering can be described in more detail if the
$J$-point, volume-averaged, correlation functions, $\bar{\xi}_{J}$,  
are extracted.
If the clustering results from the gravitational
amplification of a Gaussian primordial density field,
then the $J$-point functions are predicted to follow
a hierarchical scaling, $\bar{\xi}_{J} = S_{J} \bar{\xi}^{J-1}_{2}$.
The amplitudes $S_J$ do vary with scale, but
at a much slower rate than the volume-averaged 
correlation functions 
(Juszkiewicz, Bouchet \& Colombi 1993; Bernardeau 1994). 
This scaling behaviour has been studied extensively for cold dark matter  
in {\it N}-body simulations (e.g. Bouchet, Schaeffer \& Davis 1991; 
Baugh, Gazta\~{n}aga \& Efstathiou 1995; 
Gazta\~{n}aga \& Baugh 1995; Hivon et al. 1995; 
Colombi et al. 1996; Szapudi et al. 1999b).

Fry \& Gazta\~{n}aga (1993) proposed a simple bias model,
based on the assumption that fluctuations in the galaxy distribution
can be written as a function of the dark matter fluctuations, 
when both fields are smoothed on large scales where $\bar{\xi}_{2} \ll 1$.
The model gives predictions for the moments of the
galaxy distribution in terms of the moments for the dark matter.
To leading order in the dark matter variance, the galaxy variance
is given by $\bar{\xi}^{\rm gal}_{2} = b^{2} \bar{\xi}^{\rm DM}$,
where $b$ is usually called the linear bias.
To the same order, an additional or second order bias 
factor, $b_{2}$, is required to specify the galaxy skewness:
\begin{equation}
S^{\rm gal}_{3} = \frac{1}{b} (S^{\rm DM}_{3} + 3\frac{b_{2}}{b}). 
\label{eq:s3}
\end{equation}
Gazta\~{n}aga \& Frieman (1994) discuss the implications of the
measurements of $S_J$ from the APM Survey for the bias parameters
in this model.

In this Letter, we analyse the clustering in two optically selected redshift 
surveys that sample large volumes of the local universe. 
The Durham/UKST Survey (Ratcliffe et al. 1998) and
Stromlo-APM Survey (Loveday et al. 1996) are
magnitude limited to $b_{J} \approx 17$.
Galaxies are sparsely sampled from the parent catalogues
at a rate of 1-in-3 in the case of the Durham/UKST Survey and 1-in-20
for the Stromlo-APM Survey.
The Stromlo-APM Survey covers a three times larger solid angle than 
the Durham/UKST Survey.
By combining the results from the two surveys, the $S_{J}$ can be 
determined over a large dynamic range in cell size.

\section{COUNTS-IN-CELLS METHODOLOGY}
\label{sec:meth}

The technique of measuring the distribution of galaxy counts in cells 
is well established as a means of quantifying large scale structure 
(Peebles 1980).
The basis of the method is to throw a large number of cells onto the 
galaxy distribution in order to obtain the probability distribution of 
finding $N$ galaxies in a cell of a given size $l$.
The moments of the count probability distribution are estimated 
using the factorial moment technique, which automatically adjusts the 
moments to compensate for the sampling of a continuous density field 
using discrete galaxies (Szapudi \& Szalay 1993; Szapudi, 
Meiskin \& Nichol 1996).
The approach that we adopt here differs in two respects from most 
previous work. A similar methodology is applied to the PSCz Survey by 
Szapudi et al. (2000).

The first difference lies in how 
the higher order moments are extracted from the redshift survey.
The count probability distribution is measured in a series of volume 
limited samples of varying radial depth drawn from the flux limited survey.  
The moments obtained for a particular cell volume  
are compared between the different volume limited samples and the 
minimum variance estimate is adopted as our measurement for this scale.  
The construction of volume limited samples is straightforward: 
a maximum redshift for the sample is defined and any galaxy from the 
flux limited redshift survey that would remain visible if displaced 
out to this redshift is included in the sample 
(see, for example, Hoyle et al. 1999).

The number density of galaxies in a volume limited sample is 
effectively independent of radial distance, with small fluctuations 
due to large scale structure.
This is in direct contrast to a flux 
limited survey, where the number density changes rapidly with radius.
To analyse the count distribution in a flux limited catalogue, a weight 
must be assigned to each galaxy to compensate for the radial selection 
function. 
The analysis of volume limited samples is therefore much simpler, and 
gives equivalent results without introducing any biases: 
moreover, the task of devising an optimal weighting 
scheme and of constructing a suitable estimator of the moments to apply to 
the flux limited sample is avoided (Colombi, Szapudi \& Szalay 1998). 

This approach does, however, rely upon the assumption that 
galaxy clustering does not depend on luminosity, at least over 
the range of luminosities that we consider in our samples 
(see column 4 of Table 1 for the absolute magnitudes that define 
the volume limited samples we analyse).
Loveday et al. (1995) measured the two-point correlation function 
in redshift space for galaxies selected from the Stromlo-APM survey 
on the basis of absolute magnitude. These authors found no significant 
evidence for a difference in clustering amplitude when comparing 
samples over a much broader range of absolute magnitudes than we 
consider in our analysis.
Similar conclusions were reached by Tadros \& Efstathiou (1996) who 
analysed the amplitude of the power spectrum in different volume limited 
samples drawn from the same survey. A weak effect, at just over the  
$1 \sigma$ level, was seen only for the deepest sample, corresponding 
to an absolute magnitude of $M_{b_{J}} = -20.3$.
Hoyle et al. (1999) found that the power spectra in volume limited 
samples drawn from the Durham/UKST survey vary by less than the 
$1 \sigma$ errors as the depth of the sample is changed.
Therefore, the approximation that the intrinsic clustering in redshift space
is the same 
in different volume limited samples is fully justified by previous 
work on the surveys we analyse in this Letter.

The second difference from previous work is the treatment of 
the errors on the measured moments.
A theoretical calculation of the errors is made using 
the method described by Szapudi, Colombi \& Bernardeau (1999a)
\footnote{The FORCE package (FORtran for Cosmic Errors) was
used to compute errors. It is available upon request from its authors, 
S. Colombi (colombi@iap.fr) or IS (szapudi@cita.utoronto.ca).}. 
All the possible sources of statistical error are included in the 
calculation, namely the following.

\begin{itemize} 
\item[(i)] Finite survey volume. 
The finite volume of the survey means that fluctuations 
on scales larger than the survey volume are not probed at all.
In addition,  fluctuations on scales approaching the maximum dimensions of the 
survey are poorly sampled.
\item[(ii)] Edge effects. The density field around galaxies that lie close to 
the survey boundary is not sampled as well as it is for a galaxy that is 
well within the boundary. 
This is because cells are not permitted to straddle the survey boundary. 
\item[(iii)] Discreteness.  
The underlying density field is assumed to be continuous. 
Sampling this field discretely with galaxies makes an additional 
contribution to the measured moments. 
\item[(iv)] Sampling or measurement errors due to the finite number of cells 
used to construct the count probability distribution.
\end{itemize}

The theoretical calculation of the errors requires a number of 
quantities to be specified beforehand.
Some of these, namely the measured values of the variance 
and $S_J$ for a given cell size and the sample volume, are estimated 
directly from the sample.
The other quantities, the variance over the full sample volume 
and the higher order cumulant correlators, are treated as parameters.
The errors that we obtain are fairly insensitive to reasonable 
choices for the values of these parameters (for a full discussion 
see Szapudi et al. 1999a). 

The theoretical error calculation has been extensively tested 
for clustered distributions of dark matter using {\it N}-body 
simulations (Colombi et al. 2000).
As a further check of the calculation, we have compared the results with 
the dispersion found for the moments averaged over 40 mock Durham/UKST Survey 
samples with a redshift limit of $z = 0.06$, extracted from the Hubble 
Volume {\it N}-body simulation as described in Hoyle et al. (1999). 
For each cell size in this comparison, the measurement is selected from 
the volume limited sample that gives the minimum variance value for the 
higher order moment. 
The sample that yielded the best measurement was found to be the same 
whether the theoretical error or the dispersion over the mock 
catalogues was used. 
Furthermore, on this scale, the magnitude of the two error estimates agree 
to better than $10$ per cent.
The magnitude of the theoretical errors is within $50$ per cent of the dispersion 
over the mock catalogues on scales that do not give the minimum variance 
estimates of the higher order moments in a particular sample.

\section{RESULTS}
\label{sec:res}

\begin{figure}
{\epsfxsize=8.1truecm \epsfysize=8.1truecm 
\epsfbox[30 160 575 660]{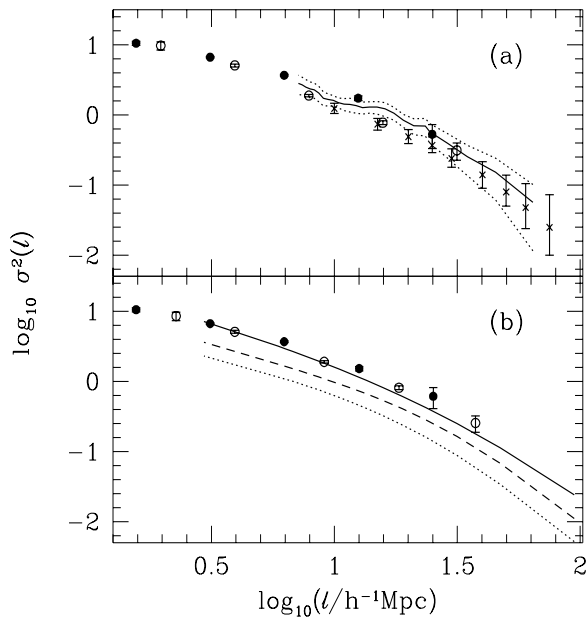}}
\caption{The variance of counts in cubical cells. In both panels, 
solid circles show the variance in the Durham/UKST Survey, 
whilst open circles the Stromlo-APM Survey results.
In panel (a), we show the variance in volume limited samples 
with $z_{\rm max}=0.06$. 
The solid line shows an estimate of the variance made from the power
spectrum measured in the same Durham/UKST sample by
Hoyle et al. (1999); the dotted lines show the 1$\sigma$ errors.
The crosses show the variance for the flux limited Stromlo-APM survey 
from Loveday et al. (1992). The error bars on the Loveday et al. points show 
$95$ per cent confidence limits. 
In panel (b), the circles show the best estimates of the variance, extracted from
a series of volume limited samples.
The lines show the variance in redshift space for the 
{\it N}-body simulations discussed in Section 4: the solid line is for a simulation 
with a linear power spectrum described by $\Gamma=0.2$ and $\sigma_{8}=1$, 
the dashed line for $\Gamma=0.5$ and $\sigma_{8}=1$ and the dotted line 
for $\Gamma=0.5$, $\sigma_{8}=0.66$ ($\sigma_{8}$ is the rms density 
fluctuation in spheres of radius $8 h^{-1}$Mpc).}
\label{fig:var}
\end{figure}

\begin{figure}
{\epsfxsize=8.4truecm \epsfysize=6.5truecm 
\epsfbox[80 260 570 574]{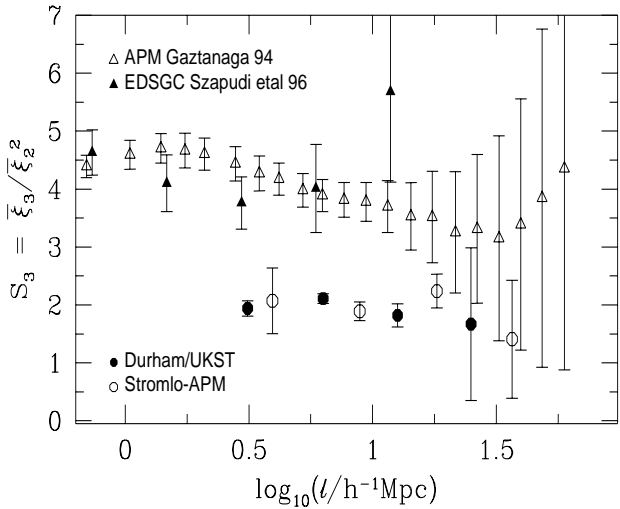}}
\caption{The skewness extracted from the redshift surveys (filled circles
show Durham/UKST results, open circles show Stromlo-APM
results) compared with the three dimensional
values inferred from the parent angular catalogues (the 
open triangles show the APM Survey results from Gazta\~{n}aga 1994, and the 
filled triangles show the results from the Edinburgh-Durham Southern Galaxy 
Catalogue from Szapudi, Meiksin \& Nichol 1996).
}
\label{fig:s3_data}
\end{figure}

The galaxy count probability distribution is measured in cubical 
cells of side $3-40 h^{-1}$Mpc in a series of volume limited samples 
drawn from the Durham/UKST and Stromlo-APM redshift surveys. 
The limiting redshifts of the samples are in the range $z \sim 0.05-0.08$, 
corresponding to maximum radial depths of $140$--$220h^{-1}$Mpc.
The higher order moments are calculated from the count probability 
distribution using the factorial moment technique introduced by 
Szapudi \& Szalay (1993).
In practice, measurement errors, (iv) in the list 
of statistical errors given in Section 2, are negligible in comparison to the 
other contributions, because on the order of $10^8$ cells are used to 
determine the count distribution at each scale. 
This massive oversampling of the density field is achieved using the 
algorithm developed by Szapudi (1998).

The second moment or variance of the galaxy distribution is shown 
in Fig.\ref{fig:var}.
In both panels, the filled circles show measurements 
obtained from the Durham/UKST Survey and the open circles show those from  
the Stromlo-APM Survey. 
Fig. \ref{fig:var}(a) shows the variance as a function of cell size 
in volume limited samples extracted from the survey, with a 
maximum redshift of  $z = 0.06$. 
Fig. 3 of Hoyle et al. (1999) shows that the number of galaxies as a function 
of the maximum redshift defining a volume limited sample peaks at this 
redshift for both surveys.  
These results are in good agreement with estimates of the variance 
made from the surveys using different techniques.
The solid line shows an independent estimate of the variance obtained from the 
the power spectrum of the same volume limited sample from the 
Durham/UKST Survey from Hoyle et al. (1999), for wavenumbers 
$k \le 0.43 h {\rm Mpc}^{-1}$. 
We have used the approximate transformation between power spectrum 
and variance given in Peacock (1991). 
The dotted lines show the 1$\sigma$ error on this estimate, which comes 
directly from the error on the measured power spectrum.
The very good level of agreement between these different 
estimates demonstrates that large volume cells genuinely 
measure fluctuations on large scales.   
Our results for a volume limited subsample of the 
Stromlo-APM survey agree well with those  obtained from the full 
magnitude limited survey shown by the crosses in Fig. \ref{fig:var}(a) 
(Loveday et al. 1992). 
The error bars on these points show the $95$ per cent percent confidence 
limits and are computed under the assumption that the distribution 
of fluctuations is Gaussian. 

\begin{figure*}
{\epsfxsize=15.1truecm \epsfysize=9.1truecm 
\epsfbox[-50 200 570 640]{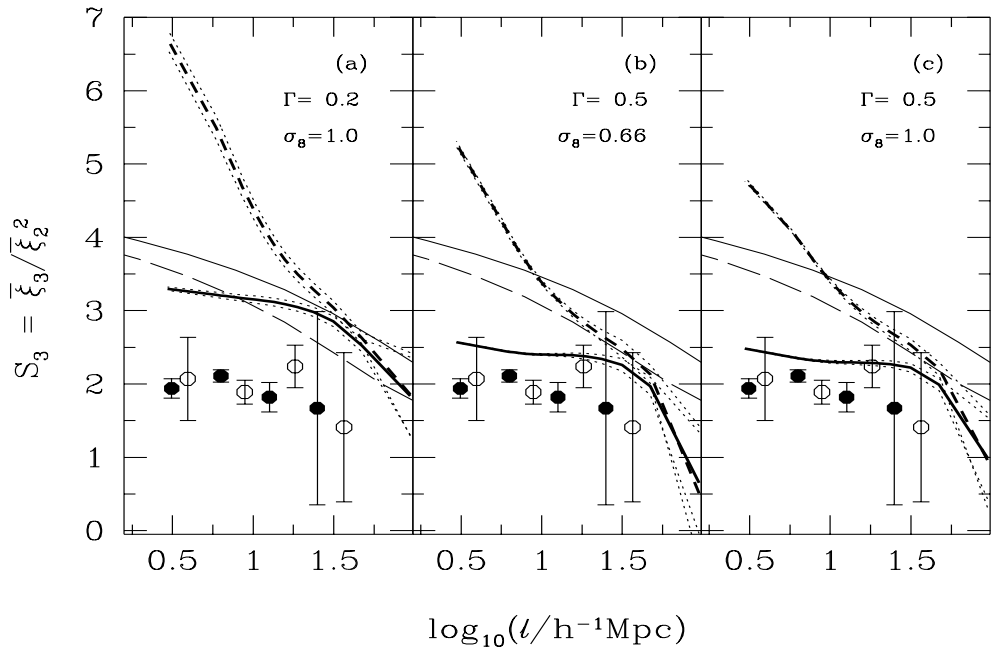}}
\caption{
A comparison of the minimum variance measurements of 
the skewness listed in Table 1 with the skewness obtained from {\it N}-body 
simulations. In each panel, the filled circles show the skewness 
measured in the Durham/UKST Survey and the open circles show 
Stromlo-APM Survey results.
The light lines show the linear perturbation theory predictions 
for $S_3$ in real space and are reproduced in each panel; 
the solid line shows the skewness for a power spectrum with $\Gamma = 0.2$, 
and the dashed line shows the result for $\Gamma=0.5$. 
The heavy lines show the simulation results and the dotted lines show the 
error on the mean over five realisations of the initial density field. 
The heavy dashed (solid) lines show the skewness measured 
in real (redshift) space.
The simulation outputs are described by the following sets of 
power spectrum parameters: (a) $\Gamma=0.2$ and $\sigma_{8}=1$, (b) 
$\Gamma=0.5$ and $\sigma_{8}=0.66$ and (c) $\Gamma=0.5$ and $\sigma_{8}=1$.
}
\label{fig:s3_sims}
\end{figure*}

In Fig. \ref{fig:var}(b), the points show the best estimates of 
the variance extracted from the two surveys as described in Section 2. 
The best estimates of the variance from the Durham/UKST survey come 
from two samples, with radial limits of $R_{\rm max}=170$$h^{-1}$Mpc and 
$R_{\rm max}=180$ $h^{-1}$Mpc; reading from left to right, 
the first two points and the last point in Fig. \ref{fig:var}(b) 
come from the $R_{\rm max}=170$$h^{-1}$Mpc sample, whilst the third and fourth 
points come from the $R_{\rm max}=180$ $h^{-1}$Mpc sample.
The smoothness of the locus traced out by the points supports our assumption
that there is no significant dependence 
of clustering strength on luminosity over the samples considered. 
The lines in \ref{fig:var}(b) 
show the variance in a set of representative CDM simulations; these 
simulations are discussed in Section 4.

The minimum variance estimates of $S_3$ from the Durham/UKST and 
Stromlo-APM surveys are listed in Table 1, along with the properties of 
the volume limited sample in which the measurement was made. 
The errors on $S_3$ are the $1\sigma$ theoretical errors predicted 
for a sample of this volume and geometry and containing the stated number 
of galaxies. 
For cubical cells between 3 and 20 $h^{-1}$Mpc, we find remarkably little 
variation in the value of $S_3$, with errors in the range $10-20$ per cent, which 
again provides further evidence against any significant luminosity 
dependence of clustering.
We obtain $S_3$ on scales larger than $20 h^{-1}$Mpc, but 
with much larger errors.

When the relative error on the estimate of $S_J$ approaches 100 per cent, 
the perturbative techniques used in the error calculation break down. 
Nevertheless, the calculation still reliably indicates that 
the errors are large and that the measurement has no significance. 
The relative errors on  $S_4$ are estimated to be
$>$100 per cent on all scales in the Stromlo-APM survey.
There is only one scale where $S_4$ can be reliably constrained from the 
Durham/UKST survey. 
This scale is also the one for which $S_3$ is most accurately
measured in this sample. 
As we expect this to be the case in general,
the values for $S_4$ on the same scale as the minimum variance measurements 
of $S_3$ are listed in Table \ref{tab:best}.
These estimates should be treated with caution as the errors are large.

\section{DISCUSSION}
\label{sec:comp}

The mean values we obtain for the skewness are in agreement with 
those found in shallower redshift surveys, though we find errors 
that are somewhat larger (e.g. Gazta\~{n}aga 1992;  
Bouchet et al. 1993; Fry \& Gazta\~{n}aga 1994; Benoist et al 1999 and 
for a comprehensive compilation of results and a more exhaustive set 
of references, see table 1 of Hui \& Gazta\~{n}aga 1999). 
Moreover, in spite of the relatively large volumes of the surveys 
considered in this Letter, we find that a significant measurement 
of $S_4$ is only possible at one scale.
There are two main reasons for the discrepancy in the magnitude of the 
estimated errors. 
The first is that some previous results are quoted 
as averages over the values of $S_3$ determined on different scales, 
exploiting the relatively flat form of $S_3$ in redshift space.
This leads to smaller errorbars under the incorrect assumption
that the individual measurements are independent.
The second reason is that not all of the contributions to 
the statistical errors listed in Section 2 were considered in previous analyses. 

We have constrained $S_3$ over a wide range of scales, 
extending beyond ${\it l}\sim20h^{-1}$Mpc, where simple models for 
bias can be tested most cleanly.
Indirect measurements of $S_3$ on these scales have been obtained  
from the {\it IRAS} 1.2-Jy Redshift Survey by fitting a parametric functional 
form for the count probability distribution to the measured 
counts (Kim \& Strauss 1998). 
The choice of function is not physically motivated and 
the error model used is simplistic and may underestimate 
the true variance (Gazta\~naga, Fosalba \& Elizalde 1999; 
Hui \& Gazta\~{n}aga 1999).     
Szapudi et al. (2000) have measured $S_3$ from the IRAS PSCz survey, 
using the same techniques employed in this paper, and find 
$S_3  = 0.87 \pm 0.48$ for cells of side $l = 37 h^{-1}$Mpc, which 
is in good agreement with the value we find, quoted in Table 1. 

We compare our measurements of $S_3$ with the values inferred from
the parent angular catalogues of the redshift
surveys in Fig \ref{fig:s3_data} (Gazta\~{n}aga 1994;
Szapudi, Meiksin \& Nichol 1996).
The results from the angular catalogues are obtained by first finding
the projected count distribution on the sky, and then applying 
a deprojection algorithm to infer the moments in three dimensions.
The algorithm requires knowledge of the survey selection function.
The deprojected angular measurements are in real space as they are
free from any distortion due to the peculiar motions of galaxies.
On large scales, the mean value we find for $S_3$ is below
that found in real space.
However, the errors are large on both measurements, and the
results are consistent at the $1 \sigma$ level.
Moreover, it is somewhat unclear exactly how important edge effects
in the angular measurements and systematic effects in the
deprojection technique are on these large scales
(Szapudi, Meiksin \& Nichol 1996; Gazta\~{n}aga \& Baugh 1998,
Gazta\~{n}aga \& Bernardeau 1998; Szapudi \& Gazta\~{n}aga 1998).

On small and intermediate scales, ${\it l} \le 15 h^{-1}$Mpc, our
determinations are below those obtained from the angular catalogues.
This is due to redshift space distortions.
The same qualitative behaviour is seen for $S_3$ measured 
in real space and redshift space in numerical simulations of hierarchical
clustering.
In Fig. \ref{fig:s3_sims}, we compare $S_3$ measured in the {\it N}-body
simulations used by Gazta\~{n}aga \& Baugh (1995), which are representative
of the behaviour in CDM models, with the redshift survey results.
The heavy dashed lines in each panel show $S_3$ in real space, and
the heavy solid lines show $S_3$ including the effects of the peculiar
motions of the dark matter.
The dotted lines show the error on the mean obtained over five realisations 
of the initial conditions (the box size of the simulations
is $378h^{-1}$Mpc).
Two different power spectra are considered: panel (a) shows a model with 
$\Gamma=0.2$ and (b) and (c) show a model with $\Gamma=0.5$ 
at two different epochs.
On large scales, the value of $S_3$ depends upon the shape of the power
spectrum and is in good agreement with the perturbation theory predictions,
which are shown by the light lines; this result was
discussed by Gazta\~{n}aga \& Baugh (1995).
The value of $S_3$ in redshift space also depends upon
the shape of the power spectrum, and is insensitive to epoch or 
equivalently to the amplitude of the fluctuations, 
as shown by Figs. \ref{fig:s3_sims}(b) and (c). 
The real and redshift space values of $S_3$ become consistent at
${\it l} \approx 20h^{-1} {\rm Mpc}$, in excellent agreement
with the comparison presented for the data in Fig. \ref{fig:s3_data}.

We now investigate how the predictions from the simulations can be reconciled 
with the observations and discuss the implications for biasing.
The model developed by Fry \& Gazta\~{n}aga (1993) predicts a relationship 
between the skewness in the galaxy distribution, $S_3^{\rm gal}$, and that 
in the underlying dark matter, $S_3^{\rm DM}$, that is applicable on 
large scales (equation \ref{eq:s3}).
The variance for the dark matter in the simulation with 
$\Gamma=0.2$ and $\sigma_{8}=1$ is very close to the observed variance 
in galaxy counts (cf. the solid line in Fig \ref{fig:var}), indicating that a relatively small 
linear bias term is required; at ${\it l} \sim 20 h^{-1}$Mpc, the linear 
bias is $b = 1.16\pm0.06 $. Furthermore, in redshift space, the linear bias 
is essentially independent of scale. 
Thus, given the scale independence of the skewness that we measure 
for galaxies and which is predicted for the dark matter from the simulations, 
we can insert the values for $S_{3}^{\rm gal}$, $S_{3}^{\rm DM}$ and $b$ into
equation \ref{eq:s3} and obtain a value for the second order bias 
term, $b_2$. 
At $l \sim 20h^{-1}$Mpc, a second order bias term of value 
$b_{2}= -0.20 \pm 0.14$ is required for the skewness of the dark matter 
to match that seen for galaxies.
For the simulation with $\Gamma = 0.5$ and $\sigma_{8}=0.66$, the linear 
bias term is larger ({cf. the dotted line in Fig. \ref{fig:var}), 
$b=1.86 \pm 0.10$, and the second order bias term is $b_{2}=1.0 \pm 0.4$. 
Hence, whilst a linear bias term is sufficient to reconcile the variance 
measured in redshift space for galaxies and for dark matter, additional 
bias terms are required to match up the results for the skewness.

A similar counts in cells analysis has been applied to the  
PSCz Survey, and yields values for $S_3$ that are in good agreement 
with those reported here at all scales (Szapudi et al. 2000).
At first sight this result is intriguing, in view of the well known 
difference in the amplitude of the two-point functions of optical 
and infra-red selected galaxies on large scales (e.g. Peacock 1997; 
Hoyle et al. 1999). 
Thus having demonstrated the need to consider a second order bias term 
in addition to the linear bias usually discussed, it would appear that 
both these quantities can depend on the way in which galaxies are selected. 
These issues are best addressed using semi-analytic models of
galaxy formation (Baugh et al. 2000).

\section*{Acknowledgments}
FH acknowledges receipt of a PPARC studentship. This work was supported 
by the PPARC rolling grant at Durham. 
We thank the Virgo Consortium (see Jenkins et al. 
1998 paper) 
for making the Hubble volume simulation 
available and for providing software to analyse the output. 
Tom Shanks and Shaun Cole provided several useful suggestions
that improved an earlier draft.
We are indebted to the efforts of all those involved in constructing 
both the parent catalogues and the redshift surveys used here, and for making 
the redshift catalogues publically available.

\end{document}